\documentclass[twocolumn,aps,floats,prb,twoside,fancyhdr,a4paper,superscriptaddress,nofootinbib,showpacs]{revtex4}
\usepackage{latexsym}
\usepackage{amsmath}
\usepackage{amssymb}
\usepackage{amsfonts}
\usepackage{graphicx}
\usepackage{dcolumn}
\usepackage{fancyhdr}
\usepackage[colorlinks]{hyperref}

\makeatletter
\def\onpage#1#2#3{\ifnum\c@page=#1 #2\else#3\fi}
\def\pagesbefore#1#2#3{\ifnum\c@page<#1 #2\else#3\fi}
\def\pagesafter#1#2#3{\ifnum\c@page>#1 #2\else#3\fi}
\makeatother

\begin{document}

\setcounter{page}{1}

\title{{\rule{16cm}{.10mm}{\bfseries{\\\vspace{3mm}SIMILARITY VERSUS SYMMETRIES.
\\ An excuse for revising a theory of time-varying ``constants".\\\rule{16cm}{.10mm}}}}}

\author{J. A. Belinch\'on}
\email{abelcal@ciccp.es}
\affiliation{Dpt. of Physics ETS Architecture. UPM Madrid. Av. Juan de Herrera N-4 28040 Espa\~na (Spain).}

\date{June 2004.}

\begin{abstract}
In this paper we compare the dimensional method with the Lie
groups tactic in order to show the limitations and advantages of
each technique.  For this purpose we study in detail a perfect
fluid cosmological model with time-varying ``constants" by using
dimensional analysis and the symmetry method. We revise our
previous conclusion about the variation of the fine structure
constant finding for example that in the radiation predominance
era if $\alpha$ varies is only due to the variation of
$e^{2}\varepsilon _{0}^{-1}$ since $c\hbar=const$ in this era.
\end{abstract}

\pacs{98.80.Hw, 04.20.Jb, 02.20.Hj, 06.20.Jr}
 \vspace{.4cm} \maketitle

\fancyhead{} \thispagestyle{fancy} \pagestyle{fancy}
\fancyhead[CE]{J.A. Belinch\'on.}
\fancyhead[CO]{\onpage{1}{}{Similarity versus Symmetries.}}
\renewcommand{\headrulewidth}{0pt}
\fancyfoot[LO,RE]{\emph{}}
\fancyfoot[RO,LE]{\emph{}}

\section{Introduction}

The main goal of this paper is to show how fruitful is to work with
dimensional techniques (DA) as compared to other more sophisticated methods as the
Lie groups (LG). For this purpose we study some cosmological models that
consider all the physical ``constants'' i.e. $G,c,\Lambda$ and $\hbar$ as
functions depending on time $t.$ In particular we will study models which
matter content is modeled by a perfect fluid. Therefore this paper is devoted
to compare the dimensional method with the Lie group tactic, by showing the
limitations and advantages of each technique.

We understand that the dimensional method is in actually a method of Lie
groups, this method has such algebraic structure, but the Pi-Theorem only
finds scaling symmetries while the Lie method finds all the possible
symmetries. One of our purposes in this paper will be to show this fact.

Cosmological models with time-varying ``constants'' have been studied for
quite some time ever since Dirac proposed a theory with a time-varying
gravitational constant $G$. Several works have investigated cosmological
models with variable cosmological constant within a framework of dissipative
thermodynamics as well as in the case of perfect fluids.

The purpose of this work is to perform a detailed study of all the possible
symmetries of a perfect fluid model with time varying constants showing that
in this case it is possible to find more solutions in addition to the scaling
one (obtained through DA). In order to carry out this study, we begin in
section 2 by outlining the equations that govern the model as well as the
notation employed. We present three models. The first of them will be
formulated without the condition $div(T_{j}^{i})=0.$ In order to make that
``constant'' $\hbar$ appears into the field equations we impose an adequate
equation of state for the energy density, in this way through this condition we
will be able to formulate a very general equation that contains to all the
``constants''. The second and third of the models verify the condition
$div(T_{j}^{i})=0.$ As in the first of our models we will consider two cases
in each model, to take a general form for the energy density and the case which verifies the equation of state $\rho=a\theta^{4}$ (black body
radiation). In order to get rid of the entropy problem in our third model we
will consider adiabatic matter creation, showing that for an adequate valour of a $\beta$-parameter
this model could be reduced to our second model. Therefore this will be our
more general model.

In section 3, we review the scaling solution obtained in previous
works by highlighting the ``assumed'' hypotheses that we need to
make in order to obtain a solution using dimensional analysis,
these are: $div(T_{j}^{i})=0,$ conservation principle, and that
the relation $G/c^{2}$ remain constant for all values of $t$
(cosmic time). In this section we will find a solution for the
field equations through two different dimensional ways. In the
first of them we apply the Pi-theorem in order to solve our most
general model, the third one, and as we will able to see this is,
in our opinion, the best method to solve the equations since it is
the simplest one and it lets us to obtain a complete solution. We
discuss some interesting relationships that arise with the
similarity method. We will revise our conclusions obtained in
previous works trying to improve them, in this way we arrive to
the conclusion that the fine structure constant varies but we
cannot know which constant or constants are the responsible of
such variation since we can only calculate the behaviour of the
constants \ $G,c,\Lambda$ and $\hbar$ but no of any
electromagnetic constant, $e,\varepsilon _{0},...$. With regard to
the second dimensional tactic we will try to solve the first of
our model, which does not verify the condition $div(T_{j}^{i})=0$.
In order to do it we shall need to impose some hypotheses to
reduce the number of unknown quatities in the field equations. We
arrive to the conclusion that the solution obtained verifies the
general conservation principle (this is a limitation of the
dimensional tactic) and that it is less general than the solution
obtained with our first dimensional method. This solution is only
valid for the case of radiation predominance i.e. $\omega=1/3$
while the above solution is valid for all value of $\omega$ that
is to say for all kind of matter and not only for radiation.

In Section 4, we work towards finding other possible solutions to
the field equations using the Lie group method. We start this
section by rewriting the field equations in such a way that we can
use the standard Lie procedure that allow us to find more
symmetries. With this tactic we only will be able the study the
second and the third model but without the possibility of
considering the case where $\rho=a\theta^{4}$ that is to say we
are not able to study the general case where $\hbar$ appears.
After outlining the equation and the constraint, we proceed to
study some cases. The first one is the obtained previously by
using dimensional analysis since dimensional analysis is just a
special class of symmetry (scaling symmetry). We would like to
emphasize that the Lie method show us that one of the assumptions
made with the dimensional method, $G/c^{2}=const.,$ is at least
correct from the mathematical point of view. This result allow us
to validate completely the solution obtained through similarity.
Nevertheless there are more symmetries, some of them correspond to
our second and third solutions

We conclude the paper by summarizing the results in section 5 and
emphasizing why the dimensional method works so well. In the
appendix we try to show how tedious is to work with the Lie method
if one tries to be rigorous.

\section{The Model\label{Model}}

In this section we will outline the field equations of three models. In the
first of them, its energy-momentum tensor does not verify the conservation
principle. The second model and the third do it but in the third one we will
consider adiabatic matter creation. In all these cases we will consider two
possibilities, a general form for the energy density $\rho $ and a specific
equation of state for this $\rho =a\theta ^{4}$, in order to introduce into
the field equations the Planck constants through the radiation ``constant'' $%
a$.

We will use in all the models the field equations in the form:
\begin{equation}
R_{ij}-\frac{1}{2}g_{ij}R=\frac{8\pi G(t)}{c^{4}(t)}T_{ij}+\Lambda (t)g_{ij},
\label{ECU1}
\end{equation}
where the energy momentum tensor is:
\begin{equation}
T_{ij}=\left( \rho +p\right) u_{i}u_{j}-pg_{ij},  \label{s2-e2}
\end{equation}
and $p=\omega \rho $ in such a way that $\omega \in \left( -1,1\right] ,$
that is to say, our universe is modeled by a perfect fluid$.$

The line element is defined by:
\begin{equation}
ds^{2}=-c^{2}dt^{2}+d\Omega ^{2},
\end{equation}
with
\begin{equation}
d\Omega ^{2}=f^{2}(t)\left[ \frac{dr^{2}}{1-kr^{2}}+r^{2}\left( d\theta
^{2}+\sin {}^{2}\theta d\phi ^{2}\right) \right],
\end{equation}
we consider ``only'' a flat model i.e. $k=0.$ as the most recent
observations suggest us\cite{K0_1}$^{-}$\cite{K0_4}:

The cosmological equations are now:
\begin{align}
2H^{\prime }+3H^{2}& =-\frac{8\pi G(t)}{c(t)^{2}}p+c(t)^{2}\Lambda (t),
\label{s2-e3} \\
3H^{2}& =\frac{8\pi G(t)}{\,c(t)^{2}}\rho +c(t)^{2}\Lambda (t),
\label{s2-e4}
\end{align}
where $H=(f^{\prime }/f)$ is the Hubble function.

Applying the covariance divergence to the second member of equation (\ref
{ECU1}) we get:
\begin{equation}
div\left( \frac{8\pi G}{c^{4}}T_{i}^{j}+\delta _{i}^{j}\Lambda \right) =0,
\label{s2-e5}
\end{equation}
that simplified is:
\begin{equation}
T_{i;j}^{j}=\left( \frac{4c_{,j}}{c}-\dfrac{G_{,j}}{G}\right) T_{i}^{j}-%
\frac{c^{4}(t)\delta _{i}^{j}\Lambda _{,j}}{8\pi G},  \label{s2-e6}
\end{equation}
which yields:
\begin{equation}
\rho ^{\prime }+3(\omega +1)\rho H=-\dfrac{\Lambda ^{\prime }c^{4}}{8\pi G}%
-\rho \dfrac{G^{\prime }}{G}+4\rho \dfrac{c^{\prime }}{c}, \label{s2-e7}
\end{equation}
where it is noted that this is the main difference with respect to
other approaches.

Therefore our \textbf{first model} is governed by the following equations:
\begin{align}
2H^{\prime }+3H^{2}& =-\frac{8\pi G(t)}{c(t)^{2}}p+c(t)^{2}\Lambda (t),
\label{mod1-1} \\
3H^{2}& =\frac{8\pi G(t)}{\,c(t)^{2}}\rho +c(t)^{2}\Lambda (t),
\label{mod1-2} \\
\rho ^{\prime }+3(\omega +1)\rho H& =-\dfrac{\Lambda ^{\prime }c^{4}}{8\pi G}%
-\rho \dfrac{G^{\prime }}{G}+4\rho \dfrac{c^{\prime }}{c}.  \label{mod1-3}
\end{align}

In order to incorporate the ``Planck constant $\hbar $'' we chose or impose
an adequate equation of state for the energy density, we use the black body
equation of state $\rho =a\theta ^{4}$ where $a=\frac{\pi ^{2}k_{B}^{4}}{%
15c^{3}\hbar ^{3}}$ so that equation (\ref{s2-e7}) is now:
\begin{equation}
4\frac{\theta ^{\prime }}{\theta }-3\left[ \frac{c^{\prime }}{c}+\frac{\hbar
^{\prime }}{\hbar }\right] +3(\omega +1)H=\frac{15\Lambda ^{\prime
}c^{7}\hbar ^{3}}{8\pi ^{3}Gk_{B}^{4}\theta ^{4}}+\frac{G^{\prime }}{G}-4%
\frac{c^{\prime }}{c},  \label{s2-e8}
\end{equation}
in this way our \textbf{first modified model} is governed by the following
equations:
\begin{widetext}
\begin{align}
2H^{\prime }+3H^{2}& =-\frac{8\pi G(t)}{c(t)^{2}}p+c(t)^{2}\Lambda (t),
\label{mod2-1} \\
3H^{2}& =\frac{8\pi G(t)}{\,c(t)^{2}}\rho +c(t)^{2}\Lambda (t),
\label{mod2-2} \\
4\frac{\theta ^{\prime }}{\theta }-3\left[ \frac{c^{\prime }}{c}+\frac{\hbar
^{\prime }}{\hbar }\right] +3(\omega +1)H& =\frac{15\Lambda ^{\prime
}c^{7}\hbar ^{3}}{8\pi ^{3}Gk_{B}^{4}\theta ^{4}}+\frac{G^{\prime }}{G}-4%
\frac{c^{\prime }}{c}.  \label{mod2-3}
\end{align}
\end{widetext}

The second class of models that we study verifies the principle
of conservation for its energy-momentum tensor i.e. we assume that $%
div(T_{j}^{i})=0,$ then equation (\ref{s2-e7}) is reduced to:
\begin{align}
\rho ^{\prime }+3(\omega +1)\rho H& =0,  \label{s2-e9} \\
-\dfrac{\Lambda ^{\prime }c^{4}}{8\pi G}-\rho \dfrac{G^{\prime }}{G}+4\rho
\dfrac{c^{\prime }}{c}& =0,  \label{s2-e10}
\end{align}
or if $\rho =a\theta ^{4},$ equation (\ref{s2-e8}) reads now:
\begin{align}
4\frac{\theta ^{\prime }}{\theta }-3\left[ \frac{c^{\prime }}{c}+\frac{\hbar
^{\prime }}{\hbar }\right] +3(\omega +1)H& =0,  \label{s2-e11} \\
\frac{15\Lambda ^{\prime }c^{7}\hbar ^{3}}{8\pi ^{3}Gk_{B}^{4}\theta ^{4}}+%
\frac{G^{\prime }}{G}-4\frac{c^{\prime }}{c}& =0.  \label{s2-e12}
\end{align}

Hence the field equations for our \textbf{second model} are (for a general
form of $\rho )$:
\begin{align}
2H^{\prime }+3H^{2}& =-\frac{8\pi G}{c^{2}}p+\Lambda c^{2},  \label{field1}
\\
3H^{2}& =\frac{8\pi G}{c^{2}}\rho +\Lambda c^{2},  \label{field2} \\
\rho ^{\prime }+3\left( \omega +1\right) \rho H& =0,  \label{field3} \\
-\frac{\Lambda ^{\prime }c^{4}}{8\pi \rho G}-\frac{G^{\prime }}{G}+4\frac{%
c^{\prime }}{c}& =0.  \label{field4}
\end{align}
and for the black body equation of state $\rho =a\theta ^{4}:$%
\begin{align}
2H^{\prime }+3H^{2}+\frac{8\pi G}{c^{2}}p-\Lambda c^{2}& =0,  \label{s2-e13}
\\
3H^{2}-\frac{8\pi G}{c^{2}}\rho -\Lambda c^{2}& =0,  \label{s2-e14} \\
4\frac{\theta ^{\prime }}{\theta }-3\left[ \frac{c^{\prime }}{c}+\frac{\hbar
^{\prime }}{\hbar }\right] +3(\omega +1)H& =0,  \label{s2-e15} \\
\frac{15\Lambda ^{\prime }c^{7}\hbar ^{3}}{8\pi ^{3}Gk_{B}^{4}\theta ^{4}}+%
\frac{G^{\prime }}{G}-4\frac{c^{\prime }}{c}& =0.  \label{s2-e16}
\end{align}

To end we study briefly the important case in which adiabatic matter creation\cite{Ma_1}$^{-}$\cite{Ma_4} can be taken into account, in order to get rid of the entropy
problem. This will be our third model. The matter creation theory is based
on an interpretation of the matter energy-stress tensor in open
thermodynamic systems, which leads to the modification of the adiabatic
energy conservation law and as a result including the irreversible matter
creation. The matter creation corresponds to an irreversible energy flow
from the gravitational field to the constituents of the particles created
and this involves the addition of a creation pressure $p_{c}$ in the matter
energy-momentum tensor which we discuss below.

The field equations that now govern our model are as follows:
\begin{align}
2H^{\prime}+3H^{2} & =-\frac{8\pi G(t)}{c^{2}(t)}(p+p_{c})+c^{2}(t)%
\Lambda(t),  \label{m1} \\
3H^{2} & =\frac{8\pi G(t)}{\,c^{2}(t)}\rho+c^{2}(t)\Lambda(t),  \label{m2} \\
n^{\prime}+3nH & =\psi,  \label{m3}
\end{align}
and taking again into account our general assumption on the conservation
principle i.e. equation (\ref{s2-e6}) with $T_{i;j}^{j}=0,$ we obtain the
two equations
\begin{align}
\rho^{\prime}+3\left( \rho+p+p_{c}\right) H & =0,  \label{maciza1} \\
\frac{\Lambda^{\prime}c^{4}}{8\pi G\rho}+\frac{G^{\prime}}{G}-4\frac
{c^{\prime}}{c} & =0,  \label{maciza2}
\end{align}
and where $n$ is the particle number density, $\psi$ is the function that
measures the matter creation, $H=f^{\prime}/f$ represents the Hubble
parameter ($f$ is the scale factor that appears in the metric), $p$ is the
thermostatic pressure, $\rho$ is energy density and $p_{c}$ is the pressure
that generates the matter creation.

The creation pressure $p_{c}$ depends on the function $\psi$. For adiabatic
matter creation this pressure takes the following form:
\begin{equation}
p_{c}=-\left[ \frac{\rho+p}{3nH}\psi\right] .  \label{w2}
\end{equation}
The state equation that we next use is the well-known expression $p=\omega
\rho,$ where $\omega=const.$ and $\omega\in(-1,1]$.

Therefore, the new set of field equations is now:
\begin{align}
2H^{\prime}+3H^{2} & =-\frac{8\pi G(t)}{c^{2}(t)}(p+p_{c})+c^{2}(t)%
\Lambda(t),  \label{ma1} \\
3H^{2} & =\frac{8\pi G(t)}{\,c^{2}(t)}\rho+c^{2}(t)\Lambda(t),  \label{ma2}
\\
\rho^{\prime}+3(\omega+1)\rho H & =(\omega+1)\rho\frac{\psi}{n},  \label{ma3}
\\
\frac{\Lambda^{\prime}c^{4}}{8\pi G\rho}+\frac{G^{\prime}}{G}-4\frac
{c^{\prime}}{c} & =0.  \label{ma4}
\end{align}

If we study from the dimensional point of view the equations (\ref{ma1}-\ref
{ma4}) it is found the following relationship between the quantities:
\begin{align}
\pi _{1}& =\frac{Gpt^{2}}{c^{2}},\text{ \ \ \ \ \ \ \ \ \ \ \ \ \ \ \ \ \ \
\ \ \ }\pi _{2}=\frac{Gp_{c}t^{2}}{c^{2}},  \notag \\
\pi _{3}& =\frac{G\rho t^{2}}{c^{2}},\text{ \ \ \ \ \ \ \ \ \ \ \ \ \ \ \ \
\ \ \ \ \ \ }\pi _{4}=c^{2}\Lambda t^{2}, \\
\pi _{5}& =\frac{n}{\psi t},\text{ \ \ \ \ \ }\pi _{6}=\frac{G^{\prime }\rho
}{\Lambda ^{\prime }c^{4}},\text{ \ \ \ \ \ }\pi _{7}=\frac{G\rho c^{\prime }%
}{\Lambda ^{\prime }c^{5}},  \notag
\end{align}
where $\pi _{1}$ and $\pi _{2}$ have been obtained from equation (\ref{ma1})
while $\pi _{3}$ and $\pi _{4}$ have been obtained from equations (\ref{ma2}%
), and $\pi _{5}$ from (\ref{ma3}). The $\pi -monomias$ $\pi _{6}$
and $\pi _{7}$ have been obtained from equation (\ref{ma4})

In this way we have the following conclusions:

\begin{enumerate}
\item  From the monomias $\pi _{1}-\pi _{3},$ we see that the quantities $%
p,p_{c}$ and $\rho $ behave in a similar way i.e.
\begin{equation}
p\thickapprox p_{c}\thickapprox \rho .  \label{r1}
\end{equation}
as we already know, since we have imposed the equation of state $%
p\thickapprox \rho ,$ but the DA of the field equations suggests us that $%
p_{c}\thickapprox \rho ,$ i.e. that the pressure $p_{c}$ behaves as the energy density.

\item  From $\pi _{4}=c^{2}\Lambda t^{2}$ we obtain a clear behaviour for
the cosmological constant:
\begin{equation}
\Lambda \thickapprox \frac{1}{c^{2}t^{2}}
\end{equation}

\item  From $\pi _{5}$ it is obtained that
\begin{equation}
\psi \thickapprox nH,  \label{r2}
\end{equation}
i.e., the DA suggest us that $\psi $ must be proportional to the
Hubble parameter. Of course other possibilities are allowed, see
for example the model presented by Prigogine and collaborators,
where $\psi \thickapprox H^{2}$.

\item  And from the monomias $\pi _{6}$ and $\pi _{7}$ we see that
\begin{equation}
\frac{G^{\prime }}{G}\thickapprox \frac{c^{\prime }}{c}.
\end{equation}
\end{enumerate}

It is noted that the DA does not understand of numerical factors,
only relations between quantities.

Since the dimensional method has suggested $\left( \psi\thickapprox
nH\right) ,$ we assume that the matter creation function follows the law
(the same than Lima et al\cite{Ma_4}:
\begin{equation}
\psi=3\beta nH,  \label{w5}
\end{equation}
where $\beta\in\left[ 0,1\right] $ is a dimensionless constant (if $\beta=0$
then there is no matter creation since $\psi=0)$. The generalized principle
of conservation $T_{i;j}^{j}=0,$ for the stress-energy tensor (\ref{maciza1}%
) leads us to:
\begin{equation}
\rho^{\prime}+3(\omega+1)\left( 1-\beta\right) \rho H=0.  \label{w4}
\end{equation}

In this way, the set of field equations for the \textbf{third model} are
now:
\begin{align}
2H^{\prime }+3H^{2}+\frac{8\pi G(t)}{c^{2}(t)}(p+p_{c})-c^{2}(t)\Lambda (t)&
=0,  \label{alg1} \\
3H^{2}-\frac{8\pi G(t)}{c^{2}(t)}\rho -c^{2}(t)\Lambda (t)& =0, \\
\rho ^{\prime }+3(\omega +1)\left( 1-\beta \right) \rho H& =0, \\
\frac{\Lambda ^{\prime }c^{4}}{8\pi G\rho }+\frac{G^{\prime }}{G}-4\frac{%
c^{\prime }}{c}& =0,  \label{alg4}
\end{align}
or equivalently for $\rho =a\theta ^{4}:$
\begin{align}
2H^{\prime }+3H^{2}+\frac{8\pi G(t)}{c^{2}(t)}(p+p_{c})-c^{2}(t)\Lambda (t)&
=0,  \label{se2-cre1} \\
3H^{2}-\frac{8\pi G(t)}{c^{2}(t)}\rho -c^{2}(t)\Lambda (t)& =0,
\label{se2-cre2} \\
4\frac{\theta ^{\prime }}{\theta }-3\left[ \frac{c^{\prime }}{c}+\frac{\hbar
^{\prime }}{\hbar }\right] +3(\omega +1)\left( 1-\beta \right) H& =0,
\label{se2-cre3} \\
\frac{15\Lambda ^{\prime }c^{7}\hbar ^{3}}{8\pi ^{3}Gk_{B}^{4}\theta ^{4}}+%
\frac{G^{\prime }}{G}-4\frac{c^{\prime }}{c}& =0.  \label{se2-cre4}
\end{align}

We must emphasize that these models with $\beta =0$ (no matter creation)
reduces to the second class of models.

Therefore these are the more general field equations. In the next
sections we will find a solution for all these field equations
through different ways beginning with the dimensional one and
ending with the Lie method.

\section{Dimensional Methods.}

In this section we will find a solution to the field equations
through two different dimensional ways\cite{ad1}$^{-}$\cite{ad6}. We will begin in the next
subsection, the simplest method, finding a solution to eqs.
(\ref{se2-cre1}-\ref {se2-cre4}) applying the Pi-theorem, while in
the following subsection, the not so simple method, we will find a
solutions to the field eqs. (\ref{mod2-1}-\ref{mod2-3}) applying a
dimensional tactic which consists in reducing the number of
variables into the equations making a simple hypotheses obtained
through DA. In this way it is obtained a simple differential
equations that admits a trivial integration.

\subsection{The simplest method.}

Our purpose in this subsection is to integrate eqs.
(\ref{se2-cre1}-\ref {se2-cre4}) that is to say to find a solution
to these equations. In this case we apply the Pi-theorem. In order
to apply the Pi-theorem in the first place we need to fix the set
of governing parameters $\wp =\wp \left(
\mathcal{C},\frak{M}\right) $ which it is composed of the set of
fundamental constants $\mathcal{C}$ and the set of fundamental quantities $%
\frak{M}$. The set of fundamental constants $\mathcal{C}$ contains
the physical constants and the characteristic constants of the
model. This model evidently
 has no physical ``constants" since these vary. Our first
characteristic constant will be obtaining by integrating equation
(\ref{w4}), that brings us to obtain the following relation
between the energy density and the scale factor and which is more
important, the constant of integration that we shall need for our
subsequent calculations:
\begin{equation}
\rho =A_{\omega ,\beta }f^{-3(\omega +1)(1-\beta )},  \label{e4}
\end{equation}
where $A_{\omega ,\beta }$ is the integration constant that
depends on the equation of state that we need to consider i.e.
constant $\omega $ and constant $\beta $ that controls the matter
creation, $\left[ A_{\omega ,\beta }\right] =L^{3(\omega
+1)(1-\beta )-1}MT^{-2}$, where we are using a dimensional base
$\frak{B=}\left\{ L,M,T\right\} ,$ see\cite{to1} for details. To
apply the Pi-theorem we need another constant. This new constants
is obtained by making a simple hypothesis about the behaviour of
the ``constants'' $G$ and $c.$ We suppose that the relation
$G/c^{2}=B,\left[ B\right] =LM^{-1}T^{0},$ remain constant in
spite of both ``constants'' vary, but in such a way that this
relationship remain constant for all $t$, the universal time.
Furthermore, in this way we are guaranteeing that it is verified
the covariance principle. With the Lie group tactic we will show
that at least this relationship has mathematical meaning since it
is deduced as solution of the field equations and not imposed as
hypothesis. Therefore the set of fundamental constants is:
$\mathcal{C=C}\left( A_{\omega ,\beta },B\right) .$ Now, the set
of fundamental quantities has only one quantity, the universal
time, $t,$ as it can be trivially deduced through the Killing
equations. Therefore $\frak{M=M}$$\left( t\right) .$

Our purpose is to show that no more hypothesis are necessary to
solve the differential equations that govern the model. Therefore
the set of governing parameters is now: $\wp =\wp \left( A_{\omega
,\beta },B,t\right) ,$ that brings us to obtain the next
relations:
\begin{eqnarray}
&&
\begin{array}{r|rrrr}
& G & A_{\omega ,\beta } & B & t \\ \hline
L & 3 & \gamma & 1 & 0 \\
M & -1 & 1 & -1 & 0 \\
T & -2 & -1 & 0 & 1
\end{array}
\notag \\
&\Longrightarrow &G\propto A_{\omega ,\beta }^{\frac{2}{\gamma +1}}B^{\frac{2%
}{\gamma +1}+1}t^{\frac{2(1-\gamma )}{\gamma +1}},
\end{eqnarray}

In this way, it can be easily obtained the rest of quantities, obtaining:
\begin{equation}
\begin{array}{l}
c\propto A_{\omega ,\beta }^{\frac{1}{\gamma +1}}B^{\frac{1}{\gamma +1}}t^{%
\frac{(1-\gamma )}{\gamma +1}}, \\
\hbar \propto A_{\omega ,\beta }^{\frac{3}{\gamma +1}}B^{\frac{3}{\gamma +1}%
-1}t^{\frac{5-\gamma }{\gamma +1}}, \\
m_{i}\propto A_{\omega ,\beta }^{\frac{1}{\gamma +1}}B^{-\frac{\gamma }{%
\gamma +1}}t^{\frac{2}{\gamma +1}}, \\
\rho \propto B^{-1}t^{-2}, \\
f\propto A_{\omega ,\beta }^{\frac{1}{\gamma +1}}B^{\frac{1}{\gamma +1}}t^{%
\frac{2}{\gamma +1}}, \\
k_{B}\theta \propto A_{\omega ,\beta }^{\frac{3}{\gamma +1}}B^{\frac{3}{%
\gamma +1}-1}t^{\frac{4-2\gamma }{\gamma +1}}, \\
a^{-1/4}s\propto A_{\omega ,\beta }^{\frac{3}{\gamma +1}}B^{\frac{3}{\gamma
+1}-\frac{3}{4}}t^{\frac{6}{\gamma +1}-\frac{3}{2}}, \\
\Lambda \propto A_{\omega ,\beta }^{\frac{-2}{\gamma +1}}B^{\frac{-2}{\gamma
+1}}t^{\frac{-4}{\gamma +1}}, \\
q=\frac{\gamma -1}{2}
\end{array}
\label{Tabla1}
\end{equation}
where $\gamma =3(\omega +1)(1-\beta )-1,$ and $q$ is the deceleration
parameter. If $\omega =1/3$ then it is observed that the entropy is not
constant
\begin{equation}
a^{-1/4}s\propto t^{\frac{3}{2(1-\beta )}-\frac{3}{2}},
\end{equation}
we can check that the next results are verified: we see that $\frac{G}{c^{2}}%
=B$ $\forall \beta $ (trivially), $\rho =a\theta ^{4}$ $\forall \beta $, $%
f=ct$ $\forall \beta $, $\Lambda \propto \frac{1}{c^{2}t^{2}}\propto f^{-2}$
while the relation $\hbar c\neq const.$ since it depends on $\beta $ (if $%
\beta =0,$ then $\hbar c=const$ but only when $\omega =1/3$ i.e.
in the radiation predominance era). We emphasize this relationship
between these two ``constants'' because as it is known from the
definition of the fine structure constants $\alpha ,$%
\begin{equation}
\alpha =\frac{e^{2}}{4\pi \varepsilon _{0}\hbar c}
\end{equation}
if $\alpha $ varies (in this epoch and only in this one) this
variation only can be caused by $e^{2}/\varepsilon _{0}.$ While in
other epoch $\hbar c\neq const$ independently of $\beta ,$ being
very difficult to explain the origin of such variation or more
exactly which one or which ones of the constant are the cause of
such variation. With this model we can only calculate the
variation of the ``constants'' $c$ and $\hbar $ (and if we are
strictly rigorous we would have to say that only for the time of
radiation predominance) and we cannot calculate the behaviour of
any electromagnetic quantity. We believe that this must be the
behaviour between this ``constants'' because as we have shown in a
previous paper\cite{to2} other relation as $\hbar \thickapprox c$
brings us to a static universe. Furthermore, the behavior obtained
here works well in the framework of the quantum cosmology as we
have pointed out in reference\cite{to3}.

In previous works we calculate the behaviour of $e^{2}\varepsilon _{0}^{-1}$
as a function of $\left( A_{\omega ,\beta },B,t\right) $ in such a way that $%
\alpha $ (dimensionless quantity) remain constant in spite of the
fact of that all ``constants'' vary but in a conspire way since
they vary but keep alpha constant. Now we believe that the
relationship, for $e^{2}\varepsilon _{0}^{-1}$ was wrong in spite
of the fact that there is a relation between $c$ and $\varepsilon _{0},$ $%
c^{2}=\left( \mu _{0}\varepsilon _{0}\right) ^{-1}$ . We cannot
obtain an expression for electromagnetic quantities only in
function of $\left( A_{\omega ,\beta },B,t\right) $ (obtained in
the framework of standard cosmology). This fact has been pointed
out in a previous work (but in a different framework since in that
work  $c=const.$) where we showed that it is impossible to
reconcile the cosmological quantities with the electromagnetic
quantities (see\cite{to4} for details). Here we have the same
situation. Therefore alpha can vary but we do not know which of
the constants are the responsible of such variation as already it
has been pointed out by M. E. Tobar\cite{tobar}.

We also can check that our model has no the so called Planck's problem since
the Planck system behaves now as:
\begin{equation}
\begin{array}{l}
l_{p}=\left( \frac{G\hbar }{c^{3}}\right) ^{1/2}\approx f(t), \\
m_{p}=\left( \frac{c\hbar }{G}\right) ^{1/2}\approx f(t), \\
t_{p}=\left( \frac{G\hbar }{c^{5}}\right) ^{1/2}\approx t,
\end{array}
\end{equation}
since the radius of the Universe $f(t)$ at Planck's epoch coincides with the
Planck's length $f(t_{p})\approx l_{p}$, while the energy density at
Planck's epoch coincides with the Planck's energy density $\rho
(t_{p})\approx \rho _{p}\approx t^{-2},$ where $\rho
_{p}=m_{p}c^{2}/l_{p}^{3}.$ See \cite{to1} for more details and the followed
method etc...

It is observed from (\ref{Tabla1}) that if we make $\beta =0$ the following
set of solutions are obtained:
\begin{equation}
\begin{array}{|l|r|r|r|r|r|r|}
\omega & 1 & 2/3 & 1/3 & 0 & -1/3 & -2/3 \\ \hline
f & 1/3 & 2/5 & 1/2 & 2/3 & 1 & 2 \\ \hline
\rho & -2 & -2 & -2 & -2 & -2 & -2 \\ \hline
\theta & -1 & -4/5 & -1/2 & 0 & 1 & 4 \\ \hline
s & -1/2 & -3/10 & 0 & 1/2 & 3/2 & 9/2 \\ \hline
G & -4/3 & -6/5 & -1 & -2/3 & 0 & 2 \\ \hline
c & -2/3 & -3/5 & -1/2 & -1/3 & 0 & 1 \\ \hline
\hbar & 0 & 1/5 & 1/2 & 1 & 2 & 5 \\ \hline
m_{i} & 1/3 & 2/5 & 1/2 & 2/3 & 1 & 2 \\ \hline
\Lambda & -2/3 & -4/5 & -1 & -4/3 & -2 & -4 \\ \hline
q & 5/2 & 3/2 & 1 & 1/2 & 0 & -1/2
\end{array}
\label{Tabla2}
\end{equation}
with: $(\omega =-1)$ corresponds to de Sitter (false vacuum) represented by
the cosmological constant (special case), $(\omega =-\frac{2}{3})$ for
domain walls, $(\omega =-\frac{1}{3})$ for strings, $(\omega =0)$ for dust
(matter predominance), $(\omega =\frac{1}{3})$ for radiation or
ultrarelativistic gases (radiation predominance), $(\omega =\frac{2}{3})$
for perfect gases, $(\omega =1)$ for ultra-stiff matter. For example, if we
take the case $\omega =0$ the table (\ref{Tabla2}) tells us that
\begin{eqnarray}
f &\propto &t^{2/3},\qquad \rho \propto t^{-2},\qquad \theta \propto
t^{0}=const.,.....  \notag \\
G &\propto &t^{-2/3},\qquad c\propto t^{-1/3},\qquad \hbar \propto t,...etc.
\end{eqnarray}

This table tells us too that if we want that our universe accelerates then
we have to impose that $-1\leq\omega<-1/3$ but we must be careful since with
this parameter we see that the temperature increases.

We can try to generalize this scenario taking into account various kinds of
matter. The idea is as follow. We can define a general energy density $%
\widetilde{\rho}$ as:
\begin{equation}
\widetilde{\rho}=\sum_{i=0}^{6}\rho_{i}
\end{equation}
where $\rho_{i}$ stands for each kind of energy density, and the parameter $%
i=0,1,...,6$ in such a way that: $i=0$ correspond to $\omega=-1$ (the false
vacuum)$,$ $i=1$ correspond to domain walls i.e. to $\omega=-2/3$, $i=2$ to $%
\omega=-1/3,$ $i=3$ to $\omega=0$, $i=4$ to $\omega=1/3,$ $i=5$ to $%
\omega=2/3$ and finally $i=6$ to $\omega=1;$ but in such a way that each
type of matter verifies the relation
\begin{equation}
p_{i}=\omega_{i}\rho_{i}
\end{equation}
in this way we define the total pressure as:
\begin{equation}
\widetilde{p}=\sum_{i=0}^{6}p_{i}
\end{equation}
but in this case we do not impose that it is verified for each kind of
matter the relation:
\begin{equation}
\rho_{i}=A_{\omega_{i}}f^{-3\left( \omega+1\right) }
\end{equation}

Our purpose is as follows: we impose that the relation $div(\widetilde{T})=0$
it is verified i.e.
\begin{equation}
\widetilde{\rho}^{\prime}+3(\widetilde{p}+\widetilde{\rho})H=0
\end{equation}
taken into account that $p_{i}=\omega_{i}\rho_{i}$ then
\begin{equation}
\widetilde{\rho}^{\prime}+3H\sum_{i=0}^{6}\left[ \left( \omega_{i}+1\right)
\rho_{i}\right] =0
\end{equation}
that we can ``\emph{approximate}'' through the relation:
\begin{equation}
\widetilde{\rho}=A_{m}f^{-m}
\end{equation}
with $m=3\sum_{i=0}^{6}\left( \omega_{i}+1\right) .$ It is proven in a
trivial way that if we consider only one type of matter we then recuperate
the above results i.e.. $m=3(\omega+1).$

Therefore with the next set of governing quantities $\frak{\wp =\wp }$ $%
(A_{m},B,t)$ we arrive to obtain the following table of results:
\begin{equation}
\begin{array}{l}
G\propto A_{m}^{\frac{2}{x+1}}B^{\frac{2}{x+1}}t^{\frac{4}{x+1}-2} \\
c\propto A_{m}^{\frac{1}{x+1}}B^{\frac{1}{x+1}}t^{^{\frac{2}{x+1}-1}} \\
\Lambda \propto A_{m}^{\frac{-2}{x+1}}B^{\frac{-2}{x+1}}t^{\frac{-4}{x+1}}
\\
\hbar \propto A_{m}^{\frac{3}{x+1}}B^{\frac{3}{x+1}-1}t^{\frac{6}{x+1}-1} \\
m_{i}\propto A_{m}^{\frac{1}{x+1}}B^{\frac{-x}{x+1}}t^{\frac{2}{x+1}} \\
f\propto A_{m}^{\frac{1}{x+1}}B^{\frac{1}{x+1}}t^{\frac{2}{x+1}} \\
\rho \propto B^{-1}t^{-2} \\
k_{B}\theta \propto A_{m}^{\frac{1}{x+1}}B^{\frac{3}{x+1}-1}t^{\frac{6}{x+1}%
-2} \\
a^{-1/4}s\propto A_{m}^{\frac{3}{x+1}}B^{\frac{3}{x+1}-\frac{3}{4}}t^{\frac{6%
}{x+1}-\frac{3}{2}} \\
q=\frac{x+1}{2}-1
\end{array}
\end{equation}
where $x+1=m=3\sum_{i=0}^{6}\left( \omega _{i}+1\right) .$ If for example we
consider an universe with dust $\left( \omega =0\right) $ and radiation $%
\left( \omega =1/3\right) $ then $m=9,$ we obtain a very surprising results
as we can see
\begin{eqnarray}
G &\propto &t^{-8/5},\quad c\propto t^{-4/5},\quad \hbar \propto
t^{-2/5},\quad \Lambda \propto t^{-2/5},  \notag \\
\quad \rho  &\propto &t^{-2},\quad f\propto t^{1/5}\quad etc...
\end{eqnarray}

As we have seen, it is obtained a complete solution of the field
equations making only one hypothesis i.e. $G/c^{2}=B.$ The method,
a direct application of the Pi-theorem, allows us to obtain, in a
trivial way, the behaviour of all the quantities under study.
Furthermore, the followed tactic allows us to generalize the model
considering a mixed of fluids (but without interactions between
them). This method is therefore, simple but powerful as we will
see in the next sections since with the rest of the tactics we
will not be able to solve this so complex model.

\subsection{Not so simple method.}

In this subsection we will try to solve the first of our models
eqs. (\ref {mod2-1}-\ref{mod2-3}), i.e. a model where all its
``constants'' are time-varying and its energy-momentum tensor does
not verify the conservation principle $ div( T_{i}^{j})\neq 0 $.
In particular we are interested in solving the eq. (\ref{mod2-3})
which contains all the information about the model. As this
equation has 6 unknown quatities, we will need to make some
hypotheses in order to simplify the original equation and to try
to integrate it. In this case we will impose some hypotheses about
the behavior between some of the quantities through DA in such a
way that these hypotheses allows us to reduce the number of
variables in the equation under study.

To solve equation (\ref{mod2-3}) we imposed two simplifying hypotheses, the
first one, that the relation $G/c^{2}=B$ (where $B$ is a $const.$) remains
constant, and the second one, that the cosmological ``constant'' verifies
the relation $\Lambda \propto \frac{d}{c^{2}t^{2}}$ with $d\in \mathbb{R}$
(this is a very strong condition)$,$ while $div( T_{i}^{j}) \neq
0 $, in this way the equation was solved perfectly. Since the constant $B$
has dimensions $\left[ B\right] =LM^{-1}$ we can get the dimensionless
monomia $\pi _{1}=\frac{\rho Bt^{2}}{b}$ where $b\in \mathbb{R}.$ With these
hypotheses and $\pi _{1}$ equation (\ref{mod2-3}) simplifies to:
\begin{widetext}
\begin{equation}
4\frac{\theta ^{\prime }}{\theta }-3\left[ \frac{c^{\prime }}{c}+\frac{\hbar
^{\prime }}{\hbar }\right] +3(\omega +1)H-\frac{15d}{4\pi ^{3}}\frac{c^{4}%
\left[ c^{\prime }t+c\right] \hbar ^{3}}{Gk_{B}^{4}\theta ^{4}t^{3}}+\frac{%
G^{\prime }}{G}-4\frac{c^{\prime }}{c}=0,  \label{p5}
\end{equation}
\end{widetext}
that has no immediate integration. We have to take into account the field
equations (\ref{mod2-2})
\begin{equation}
3H^{2}=8\pi bt^{-2}+dt^{-2},  \label{LUC1}
\end{equation}
from this we get $f=K_{\varkappa }t^{\varkappa }$ where $\varkappa =\left(
\frac{8\pi b+d}{3}\right) ^{\frac{1}{2}}$ and substituting in (\ref{p5})
together with $G=Bc^{2}$ we get
\begin{widetext}
\begin{equation}
4\frac{\theta ^{\prime }}{\theta }-3\left[ \frac{c^{\prime }}{c}+\frac{\hbar
^{\prime }}{\hbar }\right] +\frac{3(\omega +1)\varkappa }{t}-\frac{15d}{4\pi
^{3}}\frac{c^{2}\left[ c^{\prime }t+c\right] \hbar ^{3}}{Bk_{B}^{4}\theta
^{4}t^{3}}-2\frac{c^{\prime }}{c}=0,  \label{LUCIL1}
\end{equation}
\end{widetext}
i.e. one equation with $3$ unknowns.

If we want to integrate equation (\ref{LUCIL1}) we have to take a decision
on the behavior of the constant $\hbar.$

Taking $c\hbar=const.\approx\hbar=\frac{A}{c}$ then $\frac{\hbar^{\prime}}{%
\hbar}=-\frac{c^{\prime}}{c}$ yielding:
\begin{equation}
4\frac{\theta^{\prime}}{\theta}+\frac{3(\omega+1)\varkappa}{t}-\frac{15d}{%
4\pi^{3}}\frac{A^{3}\left[ c^{\prime}t+c\right] }{cBk_{B}^{4}\theta ^{4}t^{3}%
}-2\frac{c^{\prime}}{c}=0.
\end{equation}
Also if $\rho=a\theta^{4}$ and $\rho=\frac{b}{Bt^{2}}\Longrightarrow\frac
{b}{Bt^{2}}=a\theta^{4}$ we have:
\begin{equation}
k_{B}\theta=\left( \frac{15c^{3}(t)\hbar^{3}(t)b}{\pi^{2}Bt^{2}}\right) ^{%
\frac{1}{4}}=\left( \frac{15A^{3}b}{\pi^{2}B}\right) ^{\frac{1}{4}}t^{-1/2}.
\end{equation}
And substituting into the previous equation we get:
\begin{equation}
\frac{-2}{t}+\frac{3(\omega+1)\varkappa}{t}-\frac{15d}{4\pi^{3}}\frac {A^{3}%
\left[ c^{\prime}t+c\right] }{cBt^{3}\frac{15A^{3}b}{\pi^{2}Bt^{2}}}-2\frac{%
c^{\prime}}{c}=0,
\end{equation}
and simplifying
\begin{equation}
\frac{-2}{t}+\frac{3(\omega+1)\varkappa}{t}-\frac{d}{4\pi b}\left[ \frac{%
c^{\prime}}{c}+\frac{1}{t}\right] -2\frac{c^{\prime}}{c}=0,  \label{REL2}
\end{equation}
therefore we get a very simple differential equation:
\begin{equation}
\frac{c^{\prime}}{c}=\left[ \frac{12\pi b(\omega+1)\varkappa-8\pi b-d}{8\pi
b+d}\right] \frac{1}{t}
\end{equation}
integrating it we obtain easily:
\begin{equation}
c=K_{\xi}t^{\xi}
\end{equation}
where $\xi=\left[ \frac{12\pi b(\omega+1)\varkappa-8\pi b-d}{8\pi b+d}\right]
$.

We can consider another possibility. Take the group of governing quantities $%
\frak{\wp }\frak{=}\frak{\wp }\left\{ K_{\varkappa },A,t\right\} $ where $%
K_{\varkappa }$ is the proportionality constant obtained from $%
f=K_{\varkappa }t^{\varkappa }$ and $A$ is the constant establishing the
relation between $\hbar $ and $c$. The results obtained by means of the
gauge relations are:
\begin{equation}
\begin{array}{l}
G\propto K_{\varkappa }^{6}A^{-1}t^{6\varkappa -4} \\
c\propto K_{\varkappa }t^{\varkappa -1} \\
\hbar \propto K_{\varkappa }^{-1}At^{1-\varkappa } \\
k_{B}\theta \propto K_{\varkappa }^{-1}At^{-\varkappa } \\
\rho \propto K_{\varkappa }^{-4}At^{-4\varkappa } \\
m_{i}\propto K_{\varkappa }^{3}At^{-3\varkappa +2} \\
\Lambda \propto K_{\varkappa }^{-2}t^{-2\varkappa }
\end{array}
\label{OLG3}
\end{equation}
where $m_{i}$ comes from the energy density definition $\rho _{E}=\frac{%
nm_{i}c^{2}}{f^{3}}$ ($n$ stands for the particles number). We can check
that we recover the general covariance property $\frac{G}{c^{2}}=const.$ \
iif $\varkappa =\frac{1}{2}$. Similarly we can see that the following
relations are satisfied: $\rho =a\theta ^{4},$ $\rho =Af^{-4}$ (equivalent
to $div (T_{i}^{j}) =0),$ $\Lambda \propto f^{-2}$ and $f=ct$ (no
horizon problem). With the value $\varkappa =\frac{1}{2}$ we get
\begin{equation}
\begin{array}{l}
c\propto t^{-1/2},\quad \hbar \propto t^{1/2},\quad G\propto t^{-1},\quad
k_{B}\theta \propto t^{-1/2} \\
f\propto t^{1/2},\qquad \rho \propto t^{-2},\qquad m_{i}\propto t^{1/2}
\end{array}
\label{OLG4}
\end{equation}

To obtain this solution we have needed three hypotheses:

\begin{enumerate}
\item  $G/c^{2}=B,$

\item  $\Lambda \thickapprox t^{-2},$

\item  $c\hbar =const.$
\end{enumerate}

Some of them are very restrictive since they induce a scaling
solution (power law solution) and as we can see we have obtained
the same solution than in the previous subsection for a model with
$\omega =1/3$ and $\beta =0,$ i.e. with radiation predominance and
without matter creation. But, How it is possible?. We have tried
to solve eq. (\ref{mod2-3}) i.e. a model that, in principle, it
does not verify the condition $ divt( T_{i}^{j})= 0 $ however our
solution verifies such condition. As we will see in the last
section, the DA always is related to conservation principles, for
this reason in spite of working with eq. (\ref{mod2-3}) our
hypotheses obtained through DA
have taken us to solve our second model which verifies the condition $%
div( T_{i}^{j}) =0 .$

Furthermore, we must emphasize that with this tactic we have
obtained less information since our solution is only valid for the
case $\omega =1/3,$ while with the naive method, Pi-theorem, the
solution is valid for all kind of matter.

Therefore our first dimensional method, the simplest one, needs
fewer simplifying hypotheses and its solution is much more general
that the obtained one with this other dimensional method.

\section{Lie method}

As we have seen earlier, the $\pi-monomia$ is the main object in dimensional
analysis. It may be defined as a product of quantities which are invariant
under changes of fundamental units. $\pi-monomia$ are dimensionless
quantities, their dimensions are equal to unity. Dimensional analysis has
the structure of a Lie group\cite{g1}. The $\pi-monomia$ are invariant under
the action of the similarity group. On the other hand, we must mention that
the similarity group is only a special class of the mother group of all
symmetries that can be obtained using the Lie method. For this reason, when
one uses dimensional analysis, only one of the possible solutions to the
problem is obtained.

As we have been able to find a solution through dimensional analysis, it is
possible that there are other symmetries of the model, since dimensional
analysis is a reminiscent of scaling symmetries, which obviously are not the
most general form of symmetries. Hence, we shall study the model through the
method of Lie group symmetries, showing that under the assumed hypotheses
there are other solutions of the field equations. In this section we shall
show how the lie method allows us to obtain different solutions for the
field equations. In particular we seek the forms of $G$ and $c$ for which
our field equations admit symmetries i.e. are integrable (see\cite{g2}$^{-}$\cite{g11}).

An alternative use of the Lie groups have been performed by M.
Szydlowki et. al.\cite{pol1}$^{-}$\cite{pol2} where they study the Friedman
equations in order to find the correct equation of state following
pioneer works of Collins\cite{pol3}.

In order to use the Lie method, we rewrite the field equations as follows.
From (\ref{field1}) $-$ (\ref{field2}), we obtain
\begin{equation}
2\frac{f^{\prime\prime}}{f}-2\left( \frac{f^{\prime}}{f}\right) ^{2}=-\frac{%
8\pi G}{c^{2}}\left( p+\rho\right) ,
\end{equation}
and therefore
\begin{equation}
2\left( H\right) ^{\prime}=-\frac{8\pi G}{c^{2}}\left( p+\rho\right) .
\end{equation}
From equation (\ref{field3}), we can obtain
\begin{equation}
H=-\frac{\rho^{\prime}}{3\left( \left( \omega+1\right) \rho\right) },
\end{equation}
therefore
\begin{equation}
\left( \frac{\rho^{\prime}}{\rho}\right) ^{\prime}=12\pi\left(
\omega+1\right) ^{2}\frac{G}{c^{2}}\rho.
\end{equation}
Taking $12\pi\left( \omega+1\right) ^{2}=A$ and then expanding, we obtain
\begin{equation}
\rho^{\prime\prime}=\frac{\rho^{\prime2}}{\rho}+A\frac{G}{c^{2}}\rho^{2}.
\label{ber1}
\end{equation}

If we consider the case in which there is matter creation, the resulting
equation to study is now:
\begin{equation}
\rho^{\prime\prime}=\frac{\rho^{\prime2}}{\rho}+\tilde{A}\frac{G}{c^{2}}%
\rho^{2}.
\end{equation}
with $12\pi\left( \omega+1\right) ^{2}(1-\beta)^{2}=\tilde{A}.$ Therefore,
we obtain the same equation that we have obtained, namely equation (\ref
{ber1}) except the constant $\tilde{A}$, that incorporates all the
parameters that controls the matter creation.

Now, we apply the standard Lie procedure to this equation. A vector field \ $%
X$
\begin{equation}
X=\xi(t,\rho)\partial_{t}+\eta(t,\rho)\partial_{\rho},
\end{equation}
is a symmetry of (\ref{ber1}) iff
\begin{widetext}
\[
-\xi f_{t}-\eta f_{\rho}+\eta_{tt}+\left(  2\eta_{t\rho}-\xi_{tt}\right)
\rho^{\prime}+\left(  \eta_{\rho\rho}-2\xi_{t\rho}\right)  \rho^{\prime2}%
-\xi_{\rho\rho}\rho^{\prime3}+
\]%
\begin{equation}
+\left(  \eta_{\rho}-2\xi_{t}-3\rho^{\prime}\xi_{\rho}\right)  f-\left[
\eta_{t}+\left(  \eta_{\rho}-\xi_{t}\right)  \rho^{\prime}-\rho^{\prime2}%
\xi_{\rho}\right]  f_{\rho^{\prime}}=0. \label{ber2}%
\end{equation}
\end{widetext}
By expanding and separating (\ref{ber2}) with respect to powers of $%
\rho^{\prime}$, we obtain the overdetermined system:
\begin{widetext}
\begin{eqnarray}
\xi_{\rho\rho}+\rho^{-1}\xi_{\rho}  & = & 0,\label{EDP1}\\
\eta_{\rho\rho}-2\xi_{t\rho}+\rho^{-2}\eta-\rho^{-1}\eta_{\rho}  &
= & 0,\label{EDP2}\\
2\eta_{t\rho}-\xi_{tt}-3A\frac{G}{c^{2}}\rho^{2}\xi_{\rho}-2\rho^{-1}\eta_{t}
& = & 0,\label{EDP3}\\
\eta_{tt}-A\left(  \frac{G^{\prime}}{c^{2}}-2G\frac{c^{\prime}}{c^{3}}\right)
\rho^{2}\xi-2\eta A\frac{G}{c^{2}}\rho+\left(  \eta_{\rho}-2\xi_{t}\right)
A\frac{G}{c^{2}}\rho^{2}  & = & 0. \label{EDP4}%
\end{eqnarray}
\end{widetext}

Solving (\ref{EDP1}-\ref{EDP4}), we find that
\begin{equation}
\xi(t,\rho)=-2et+a,\quad\eta(t,\rho)=\left( bt+d\right) \rho,  \label{ber3}
\end{equation}
subject to the constrain
\begin{equation}
\frac{G^{\prime}}{G}=2\frac{c^{\prime}}{c}+\frac{bt+d-4e}{2et-a},
\label{ber4}
\end{equation}
with $a,b,e,$ and $d$ as constants. In order to solve (\ref{ber4}), we
consider the following cases.

\subsection{\textbf{Case I:} $b=0$ and $d-4e=0$}

\label{sol1}

In this case, the solution (\ref{ber4}) reduces to
\begin{equation}
\frac{G^{\prime}}{G}=2\frac{c^{\prime}}{c}\Longrightarrow\frac{G}{c^{2}}%
=B=const.
\end{equation}
which means that ``constants'' $G$ and $c$ vary but in such a way that the
relation $\frac{G}{c^{2}}$ remains constant.

The solution obtained through Dimensional Analysis needs to make this
relations as hypothesis in order to obtain a complete solution for the field
equations. This case shows us that such hypothesis is correct (at least has
mathematical sense).

The knowledge of one symmetry $X$ might suggest the form of a particular
solution as an invariant of the operator $X$ i.e. the solution of
\begin{equation}
\frac{dt}{\xi\left( t,\rho\right) }=\frac{d\rho}{\eta\left( t,\rho\right) },
\label{rho}
\end{equation}
this particular solution is known as an invariant solution (generalization
of similarity solution), therefore the energy density is obtained as
\begin{equation}
\frac{dt}{-2et+a}=\frac{d\rho}{4e\rho}\Longrightarrow\rho=\frac{1}{\left(
2et-a\right) ^{2}},
\end{equation}
for simplicity we adopt
\begin{equation}
\rho=\rho_{0}t^{-2},
\end{equation}

Once we have obtained $\rho$, we can obtain $f$ (the scale factor) from
\begin{equation}
\rho=A_{\omega}f^{-3\left( \omega+1\right) }\Longrightarrow f=\left(
A_{\omega}t\right) ^{\frac{2}{3\left( \omega+1\right) }},  \label{f}
\end{equation}
in this way we find $H$ and from eq. (\ref{field2}), we obtain the behaviour
of $\Lambda$ as:
\begin{equation}
c^{2}\Lambda=3H^{2}-\frac{8\pi G}{c^{2}}\rho,  \label{lamda}
\end{equation}
and therefore,
\begin{equation}
\Lambda=\left( 3\beta^{2}-8\pi B\rho_{0}\right) \frac{1}{c^{2}t^{2}}=\frac{l%
}{c^{2}t^{2}}.
\end{equation}
If we replace all these results into eq. (\ref{field4}), then we obtain the
exact behaviour for $c,$ i.e.,
\begin{equation}
-\left( \frac{1}{t}+\frac{c^{\prime}}{c}\right) \lambda=\frac{c^{\prime}}{c},
\end{equation}
where $\lambda=\frac{l}{8\pi B\rho_{0}}$, with $\lambda\in\mathbb{R}^{+}$,
i.e. is a positive real number and thus,
\begin{equation}
c=c_{0}t^{-\alpha},
\end{equation}
with $\alpha=\left( \frac{\lambda}{1+\lambda}\right) .$

\begin{figure*}[]
\includegraphics[height=1.612in,width=5.3454in]{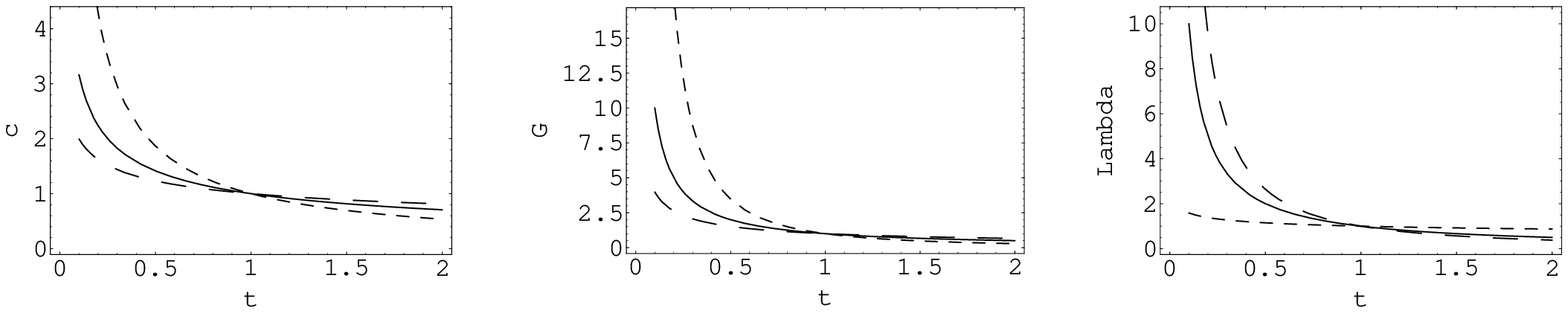} %
\caption{We see the behaviour of $G,c$ and $\Lambda$ for the first class of
solutions for different values of $\protect\alpha:\protect\alpha=0.5$ (solid
curve), $\protect\alpha=0.9$ (dotted curve), $\protect\alpha=0.3$ (matter
era)(dashed curve). In all cases the constants are decreasing functions.}
\label{fig1}
\end{figure*}

Hence, in this case we have found that (see fig.\ref{fig1}):
\begin{eqnarray}
G&=&G_{0}t^{-2\alpha},c=c_{0}t^{-\alpha},\Lambda
=\Lambda_{0}t^{-2(1-\alpha)},  \notag \\
f&=&\left( A_{\omega}t\right) ^{\frac{2}{3\left( \omega+1\right) }%
},\rho=\rho_{0}t^{-2}.
\end{eqnarray}
This is the solution that we have obtained with dimensional analysis in the
previous section. 

\subsection{\label{sol2}\textbf{Case II}, $b=a=0$}


In this case, we find that
\begin{equation}
\frac{G}{c^{2}}=\widetilde{B}t^{\varkappa},
\end{equation}
where $\varkappa=\delta-2$ and $\delta=\frac{d}{2e}.$ On following the same
procedure as above, we find that

\begin{equation}
\frac{dt}{\xi}=\frac{d\rho}{\eta}\Longrightarrow\rho=\rho_{0}t^{-\delta},
\end{equation}
we must impose the condition $sign(d)=sign(e)$, i.e., $\delta\in\mathbb{R}%
^{+},$ in order that the solution has some physical meaning that the energy
density is a decreasing function of time $t$. It is observed that if $d=4e$
then we obtain same solution that the obtained one in the case I. The scale
factor is found to be
\begin{equation}
f=K_{f}t^{\frac{\delta}{3\left( \omega+1\right) }},
\end{equation}
where $K_{f}$ is an integration constant, and therefore, the Hubble
parameter is:
\begin{equation}
H=\frac{\delta}{3\left( \omega+1\right) t},
\end{equation}
which is similar to the scale factor obtained in case I. To obtain the
behaviour of the ``constants'' $G$, $c$ and $\Lambda$, we follow the same
steps as in case I, i.e., from
\begin{equation}
c^{2}\Lambda=3H^{2}-\frac{8\pi G}{c^{2}}\rho,
\end{equation}
we obtain the behaviour of $\Lambda$ being:
\begin{equation}
\Lambda=\frac{l}{c^{2}t^{2}},
\end{equation}
where, $l=\left( K_{1}-K_{2}\right)$, $K_{1}=\frac{\delta^{2}}{3\left(
\omega+1\right) ^{2}}$ and $K_{2}=8\pi\rho_{0}\widetilde{B}$ i.e., $l\in%
\mathbb{R}^{+}$. Therefore,
\begin{equation}
\Lambda^{\prime}=-\frac{2l}{c^{2}t^{2}}\left( \frac{c^{\prime}}{c}+\frac
{1}{t}\right)
\end{equation}

If we substitute all this results into the next equation
\begin{equation}
\frac{\Lambda^{\prime}c^{4}}{8\pi G\rho}+\frac{G^{\prime}}{G}-4\frac
{c^{\prime}}{c}=0,
\end{equation}
we obtain an ODE for $c$, i.e.,
\begin{equation}
\frac{c^{\prime}}{c}\left( \lambda-2\right) =-\left( \lambda-2+\delta\right)%
\frac{1}{t}
\end{equation}
where, $\lambda=\left( -\frac{l}{4\pi\rho_{0}\widetilde{B}}\right)$, $%
\lambda\in\mathbb{R}^{-}$, which leads to
\begin{equation}
c=c_{0}t^{-\alpha}
\end{equation}
with $\alpha=\left( 1+\frac{\delta}{\lambda-2}\right) $ such that $%
\alpha\in[0,1)$. In this way we can find the rest of quantities:
\begin{equation}
G=G_{0}t^{-2\left( \alpha+1\right) +\delta },\quad{\ \ \ \ }%
\Lambda=\Lambda_{0}t^{-2\left( 1-\alpha\right) },
\end{equation}
note that $\alpha<1$. The case $\alpha=1\Longleftrightarrow\delta=0$ is
forbiden and $\alpha=0$ brings us to the limiting case of the $G,\Lambda$
variable cosmologies$^{41}$.

\begin{figure}[h!]
\begin{center}
\includegraphics[height=1.5835in,width=2.5036in]{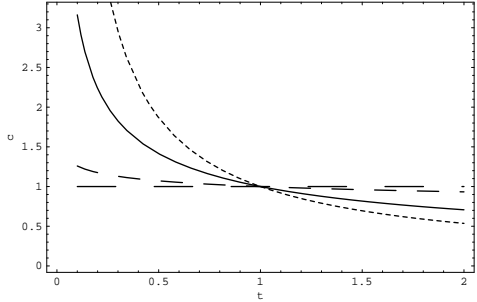}
\end{center}
\caption{Time variation of $c(t)$ for the second class of solutions for
different values of $\protect\alpha:\protect\alpha=0.5$ (solid curve), $%
\protect\alpha=0.9$ (dotted curve), $\protect\alpha=0.1$ (dashed curve) and $%
\protect\alpha=0.000001$ (long dashed curve), the last solution describes
the case $c(t)=const.$}
\label{fig2}
\end{figure}

\begin{figure}[h]
\begin{center}
\includegraphics[height=1.5852in,width=2.5018in]{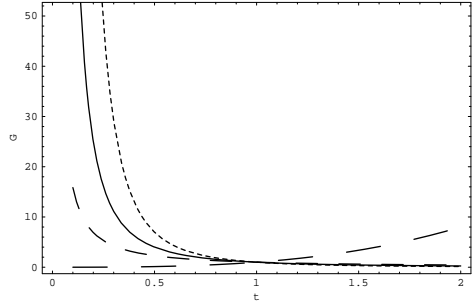}
\end{center}
\caption{The variation of the gravitational ``constant'' $G(t)$, for
different values of $\protect\alpha$ and $\protect\delta.$ $:\protect\alpha%
=0.5$ and $\protect\delta=1$ (solid curve), $\protect\alpha=0.9$ and $%
\protect\delta=1$ (dotted curve), $\protect\alpha=0.1$ and $\protect\delta=1$
(dashed curve) and $\protect\alpha=0.000001$ and $\protect\delta=5$ (long
dashed curve), the last curve describes a growing solution.}
\label{fig3}
\end{figure}

\begin{figure}[h]
\begin{center}
\includegraphics[height=1.5844in,width=2.5074in]{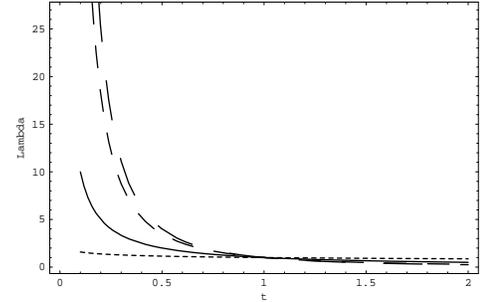}
\end{center}
\caption{Time variation of $\Lambda(t)$ for the second class of solutions
for different values of $\protect\alpha:\protect\alpha=0.5$ (solid curve), $%
\protect\alpha=0.9$ (dotted curve), $\protect\alpha=0.1$ (dashed curve) and $%
\protect\alpha=0.000001$ (long dashed curve). In all cases, $\Lambda(t)$ is
a decreasing function.}
\label{fig4}
\end{figure}

We notice that this solution is very similar to the case I but in this case
all the parameters are perturbed by $\delta$ and more important is the
result, $\frac{G}{c^{2}}=\widetilde{B}t^{\varkappa}$ (see figs.\ref{fig2},
\ref{fig3} and \ref{fig4}).

\subsection{\textbf{Case III}, $b=e=0$}

\label{sol3}

Following the same procedure as above, we find in this case that such
restrictions imply $\xi(t,\rho)=a,$ $\eta(t,\rho)=d\rho$ and therefore:
\begin{equation}
\frac{G^{\prime}}{G}=2\frac{c^{\prime}}{c}-\frac{d}{a},
\end{equation}
which brings us to:
\begin{equation}
\frac{G}{c^{2}}=K\exp(-\alpha t),
\end{equation}
where $\frac{d}{a}=\alpha$ and note that $\left[ K\right] = \left[ B\right] $
i.e has the same dimensional equation,
\begin{equation}
\frac{dt}{\xi\left( t,\rho\right) }=\frac{d\rho}{\eta\left( t,\rho\right) }%
\Longrightarrow\frac{dt}{a}=\frac{d\rho}{d\rho}\Longrightarrow\rho=\rho
_{0}\exp\left( \alpha t\right) ,
\end{equation}
this expression only has sense if $\alpha\in\mathbb{R}^{-},$ note that $%
\left[ \alpha\right] =T^{-1}.$

The scale factor $f$ satisfies the relationship:
\begin{equation}
\rho=A_{\omega}f^{-3\left( \omega+1\right) }\Longrightarrow
f=K_{f}\exp\left( \alpha t\right) ^{\frac{-1}{3\left( \omega+1\right) }},
\end{equation}
that is to say, it is a growing function without singularity. In this way,
we find that
\begin{equation}
H=-\frac{\alpha}{3\left( \omega+1\right)}=conts.\quad\quad H>0.
\end{equation}
The cosmological ``constant'' is obtained as
\begin{equation}
c^{2}\Lambda=\frac{\alpha^{2}}{3\left( \omega+1\right) ^{2}}-8\pi K\rho
_{0}\Longrightarrow c^{2}\Lambda=l,
\end{equation}
note that $\left[ l\right] =T^{-2},$ if we replace all these results into
eq. (\ref{field4}) then we shall obtain the exact behaviour for $c,$ i.e.
\begin{equation}
\left( \frac{l}{8\pi K\rho_{0}}+2\right) \frac{c^{\prime}}{c}=\alpha,
\end{equation}
and hence,
\begin{equation}
c=K\exp(c_{0}t),
\end{equation}
where $c_{0}=\frac{\alpha}{\left( \frac{l}{8\pi K\rho_{0}}+2\right) }$ with $%
c_{0}\in\mathbb{R}^{-}$ since $\alpha\in\mathbb{R}^{-}$, that is, $c$ is a
decreasing function on time $t.$

\begin{figure*}[]
\includegraphics[height=2.1871in,width=6.1445in]{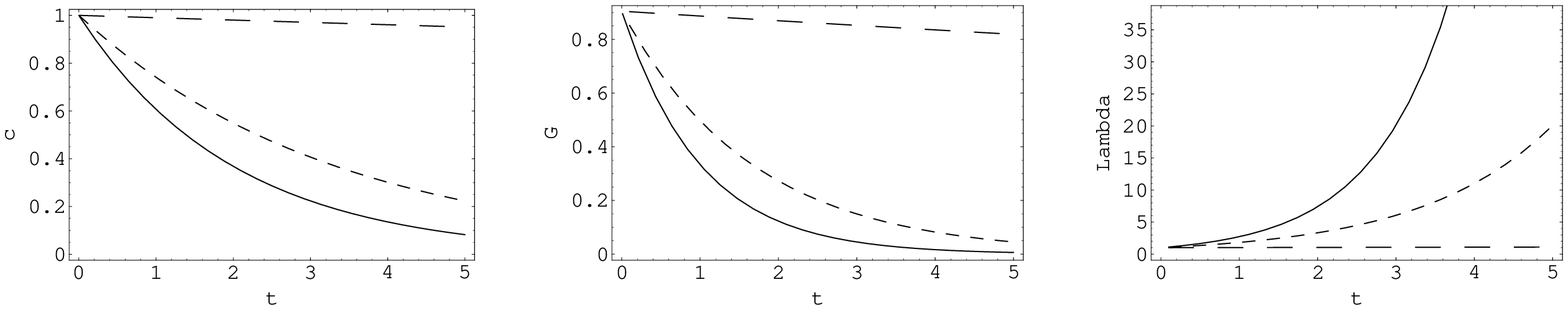}%
\caption{We see the behaviour of constants for the third class of solutions
for different values of $c_{0}:c_{0}=-0.5$ (solid curve), $c_{0}=-0.3$
(dotted curve), $c_{0}=-0.01$ (dashed curve) and $\protect\alpha=-0.1$. In
all cases, $\Lambda(t)$ is a growing function.}
\label{fig5}
\end{figure*}

In this case, we have found
\begin{eqnarray}
c &=&K\exp(c_{0}t), \\
G &=&G_{0}\exp(\left( -\alpha+2c_{0}\right) t), \\
\Lambda &=&l\exp(c_{0}t)^{-2},
\end{eqnarray}
therefore the solutions for this case are (see fig. \ref{fig5}):
\begin{eqnarray}
G &=&G_{0}\exp(\left( -\alpha+2c_{0}\right) t), c=K\exp (c_{0}t),  \notag \\
\Lambda &=&l\exp(c_{0}t)^{-2}, \rho =\rho_{0}\exp\left( \alpha t\right),
\notag \\
f &=&K_{f}\exp\left( \alpha t\right) ^{\frac{-1}{3\left( \omega+1\right) }}.
\end{eqnarray}

We have only obtained three solutions but playing with the
constants $a,b,d$ and $e$ more solutions can be obtained (without
forgetting that we are only interested in solutions with physical
sense). As we indicated in the introduction and throughout all the
paper, we have obtained in the first of our solutions (Case I),
the same solution that the obtained one with the dimensional
methods. In this occasion we have not needed to make a previous
hypothesis in order to arrive a complete solution, this is
therefore one of the advantages of this method as opposed to the
dimensional one. We have also studied two other cases which can be
considered as physically relevant solutions since $f$ is a growing
function on time and $\rho$ is a decreasing function on time. They
could describe very early cosmological solutions (inflationary
ones).

We see that with this sophisticated method we can only solve the
models 2 and 3 in which the energy density is expressed in a
generic form $i.e.$ we cannot consider the complicated case in
which appears the Planck's constant $\hbar $ (the disadvantage).
Although we have not been excessively scrupulous (formal in the
procedure (see appendix)) the method becomes more and more
complicated in function of the complexity of the equation under
study.

\section{Why the dimensional method is so wonderful and conclusions.}

As have been pointed out by Carr and Coley\cite{h1}, the existence of self-
similar solutions (Barenblatt and Zeldovich\cite{h2}) is related to
conservation laws and to the invariance of the problem with respect to the
group of similarity transformations of quantities with independent
dimensions. This can be characterized within general relativity by the
existence of a homotetic vector field and for this reason one must
distinguish between geometrical and physical self-similarity. Geometrical
similarity is a property of the spacetime metric, whereas physical
similarity is a property of the matter fields (our case). In the case of
perfect fluid solutions admitting a homotetic vector, geometrical
self-similarity implies physical self-similarity.

As we show in this section as well as in previous works, the assumption of
self-similarity reduces the mathematical complexity of the governing
differential equations. This makes such solutions easier to study
mathematically. Indeed self-similarity in the broadest Lie sense refers to
an invariance which allows such a reduction.

Perfect fluid space-times admitting a homotetic vector within general
relativity have been studied by Eardley\cite{h3}. In such space-times, all
physical transformations occur according to their respective dimensions, in
such a way that geometric and physical self-similarity coincide. It is said
that these space-times admit a transitive similarity group and space-times
admitting a non-trivial similarity group are called self-similar. Our model
i.e. a flat FRW model with a perfect fluid stress-energy tensor has this
property and as already have been pointed out by Wainwright\cite{h4}, this
model has a power law solution.

Under the action of a similarity group, each physical quantity $\phi$
transforms according to its dimension $q$ under the scale transformation.
For space-times with a transitive similarity group, dimensionless quantities
are therefore spacetime constants. This implies that the ratio of the
pressure of the energy density is constant so that the only possible
equation of state is the usual one in cosmology i.e. $p=\omega\rho$, where $%
\omega$ is a constant. In the same way, the existence of homotetic vector
implies the existence of conserved quantities.

In this paper we have studied the behaviours of time-varying ``constants'' $%
G,c$ and $\Lambda$ in a perfect fluid model. We began reviewing the scaling
solution obtained through dimensional analysis.

To obtain this solution, we imposed the assumption, $div(T_{j}^{i})=0$, from
which we obtained the dimensional constant $A_{\omega}$ that relates $%
\rho\propto f^{-3(\omega+1)}$ and the relationship
$G/c^{2}=const.=B$ remaining constants for all value of $t,$ i.e.
$G$ and $c$ vary but in such a way that $G/c^{2}$ remain constant.
With these two hypothesis, we have obtained a scaling solution for
all the quantities. In this context, the solution obtained through
dimensional analysis show us that the ``constants'' $G,c$ and
$\Lambda$ are decreasing functions of time, but in this case
decrease slowly than in the radiation predominance era, while
$\rho$ and $f$ behave as in the FRW model solving the horizon
problem.

Therefore the DA is very simple and allows us to study very complex models.
Nevertheless it has very serious limitations, we are restricted to models that
verify the conservation principle $div(T_{i}^{j})=0$ and work only with flat
models i.e. $k=0,$ for models with $k\neq0$ there are not scaling solutions.

Since we have been able to found a solution through similarity, i.e. through
dimensional analysis, it is possible that there are other symmetries of the
model, since dimensional analysis is a reminiscent of scaling symmetries,
which obviously are not the most general form of symmetries. Therefore, we
studied the model through the method of Lie group symmetries, showing that
under the assumed hypotheses, there are other solutions of the field
equations (an advantage of this method).

The first solution obtained is the already obtained one through similarity,
but, in this case we have showed the condition $G/c^{2}$ arises as a result
and not as an ad-hoc condition. We also have studied two other cases which
can be considered as physically relevant solutions since $f$ is a growing
function on time and $\rho$ is a decreasing function on time. They could
describe very early cosmological solutions (inflationary ones).

The lie method maybe is the most powerful but has drawbacks, it is
very complicate. We cannot, we do not know, solve the most general
of our models, but it allows us to find more solutions than with
de DA method and without the necessity of making previous
hypotheses

\begin{acknowledgments}
I wish to acknowledge to Javier Aceves his translation into
English of this paper.
\end{acknowledgments}

\appendix
\section{Complete Lie Method for the Case I (\ref{sol1})}

In this section we try to show how the Lie method works in order to obtain a
complete solution to equation
\begin{equation}
\rho^{\prime\prime}=\frac{\rho^{\prime2}}{\rho}+K\rho^{2},  \label{A_ber1}
\end{equation}
taking $K=AB$ with $12\pi\left( \omega+1\right) ^{2}=A$ and $%
B=G/c^{2}=const. $ We will follow the ways of the \emph{canonical coordinates%
}.

The first step is to determine the admissible algebra $L_{2}.$ As in section
(\ref{sol1}) the standard procedure brings us to obtain the overdetermined
system:
\begin{align}
\xi_{\rho\rho}+\rho^{-1}\xi_{\rho} & =0,  \label{A_EDP1} \\
\eta_{\rho\rho}-2\xi_{t\rho}+\rho^{-2}\eta-\rho^{-1}\eta_{\rho} & =0,
\label{A_EDP2} \\
2\eta_{t\rho}-\xi_{tt}-3K\rho^{2}\xi_{\rho}-2\rho^{-1}\eta_{t} & =0,
\label{A_EDP3} \\
\eta_{tt}-2\eta K\rho+\left( \eta_{\rho}-2\xi_{t}\right) K\rho^{2} & =0.
\label{A_EDP4}
\end{align}

Solving (\ref{A_EDP1}-\ref{A_EDP4}), we find that
\begin{equation}
\xi(t,\rho)=b+at,\qquad\eta(t,\rho)=-2a\rho,  \label{ber3}
\end{equation}
where $a$ is a constant. Thus equation (\ref{A_ber1}) admits two linearly
independent operators
\begin{equation}
X_{1}=\partial_{t},\text{ \ \ \ \ \ }X_{2}=t\partial_{t}-2\rho\partial_{%
\rho}.  \label{a_operator}
\end{equation}

The second step consists of determining the type of algebra that form these
two operators. We see that
\begin{equation}
\left[ X_{1},X_{2}\right] =X_{1}\text{ \ \ with }\xi_{1}\eta_{2}-\xi_{2}%
\eta_{1}=-2\rho\neq0,
\end{equation}
therefore $X_{1}$ and $X_{2}$ spand a solvable non abelian algebra type III
(see Ibragimov (1999)\cite{g4} for details, theorem 12.6. page. 287).

The next step consists in determining the integrating change of variable.
Upon introducing the canonical variables for $X_{1}$ $\left(
X_{1}s=0,X_{1}u=1\right) $ given by
\begin{equation}
s=\rho\text{ \ \ and \ \ }u(s)=t,  \label{A_cv}
\end{equation}
and therefore
\begin{equation}
t=u(s)\text{ \ \ \ \ \ }\rho=s.
\end{equation}

We transform the operator $X_{1}$ and $X_{2}$ to the form:
\begin{equation}
X_{1}=\partial_{u},\text{ \ \ \ \ \ }X_{2}=-2s\partial_{s}+u\partial_{u}.
\end{equation}
their difference from the corresponding operators of type III by the factor $%
-2$ in $X_{2}$ does no hinder the integration.

We rewrite equation (\ref{A_ber1}) in the new variables (\ref{A_cv})
obtaining:
\begin{equation}
u^{\prime\prime}=-\frac{\left( 1+Ks^{3}u^{\prime2}\right) u^{\prime}}{s},
\end{equation}
the canonical variables allow us to reduce the order of our ode. If we
follow this tactic, then it is obtained the next first order ode (type
Bernoulli),
\begin{equation}
z^{\prime}=-\frac{z}{s}-Kz^{3}s^{2},  \label{bernoulli}
\end{equation}
where $s=\rho$ and $z=du/ds=(1/\rho^{\prime})$ and which solution is:
\begin{equation}
z(s)=\pm\frac{1}{s\sqrt{2Ks+C_{1}}}.  \label{A_sol_u}
\end{equation}

The last step consists in obtaining the solution in the original variables,
being this:
\begin{equation}
\rho ^{\prime } =\rho \sqrt{2K\rho +C_{1}}
\end{equation}
this last ode has the following solution
\begin{equation}
t+\frac{2\arctan \left( \sqrt{\frac{2\rho +C_{1}}{C_{1}}}\right) }{\sqrt{%
C_{1}}}+C_{2}=0
\end{equation}
and therefore
\begin{equation}
\rho =\frac{1+\tan \left( \frac{t+C_{2}}{2C_{1}}\right) ^{2}}{2KC_{1}^{2}}.
\end{equation}
Therefore this is the most general solution to equation (\ref{A_ber1}) but
as it can be observed this seems to be non-physical, for this reason we will
seek more change of variables in order to find a better (physical) solutions.

We would like to emphasize how the Lie procedure brings us to
solve equation (\ref{bernoulli}). Of course this ode can be
trivially integrated but, at this point we prefer to integrate it
following two tedious ways. In the first of them we study equation
(\ref{bernoulli}) through the Lie method i.e. following all the
procedure shown above and with the second method we try to find
and integration factor (Lie integration factor) which brings us to
a direct integration of our equation.

Therefore studying equation (\ref{bernoulli}) through the Lie method it is
obtained the overdetermined system:
\begin{equation*}
\eta _{s}+\left( \eta _{z}-\xi _{s}\right) \left( -\frac{z}{s}%
-Kz^{3}s^{2}\right) -\xi _{z}\left( -\frac{z}{s}-Kz^{3}s^{2}\right) ^{2}-
\end{equation*}
\begin{equation}
-\xi \left( \frac{z}{s^{2}}-2Kz^{3}s\right) -\eta \left( -\frac{1}{s}%
-3Kz^{2}s^{2}\right) =0  \label{edp}
\end{equation}

Solving (\ref{edp}), we find that equation (\ref{bernoulli}) admits the
following linearly independent operators
\begin{align}
X_{1} & =z^{3}s^{2}\partial_{z},\quad X_{2}=\left( z-2Kz^{3}s^{3}\right)
\partial_{z},  \notag \\
X_{3} & =\partial_{s}-\frac{z}{s}\partial_{z},\quad X_{4}=s\partial_{s}-%
\frac{3}{2}z\partial_{z}.
\end{align}

With one of these fields, we obtain a new set of variables which brings us
to obtain a new ode which will be trivially integrated. For example, from $%
X_{3}$ it is obtained
\begin{equation}
i=z(s)s\text{, \ \ \ \ and \ \ \ \ }h(i)=s,
\end{equation}
and the inverse transformation is:
\begin{equation}
s=h(i)\text{ \ \ \ and \ \ \ }z(s)=\frac{i}{h(i)},
\end{equation}
in these new variables equation (\ref{bernoulli}) yields
\begin{equation}
h^{\prime}=\frac{-1}{Ki^{3}},
\end{equation}
which is integrated by quadratures
\begin{equation}
h=\frac{1}{2Ki^{2}}+C_{1}
\end{equation}

Now we obtain the solution of the equation in the original variables,
finding that
\begin{equation}
z=\pm\frac{1}{s\sqrt{2Ks+C_{1}}}
\end{equation}

Our second way consists in obtaining a Lie integration factor $\mu $ (for
our ode (\ref{bernoulli})) in such a way that $\mu \ast ode$ is an exact
equation. We find the following integration factor:
\begin{equation}
\mu =\frac{1}{z^{3}s^{2}},
\end{equation}
which bring us to the following solution
\begin{equation}
z=\pm \frac{1}{s\sqrt{2Ks+C_{1}}}.
\end{equation}

Now we see other changes of variables induced by other operators. If for
example we calculate the canonical variables through the second operator (%
\ref{a_operator}) $X_{2}=t\partial_{t}-2\rho\partial_{\rho}$ then
\begin{equation}
s=\rho t^{2}\text{ \ \ and \ \ }u(s)=\ln t,
\end{equation}
and therefore
\begin{equation}
t=e^{u(s)}\text{ \ \ \ \ and \ \ \ \ }\rho=\frac{1}{e^{2u(s)}},
\end{equation}
with these new variables, the original equation yields
\begin{equation}
u^{\prime\prime}=-\frac{u^{\prime}\left( 1+su^{\prime}+\left( Ks-2\right)
s^{2}u^{\prime2}\right) }{s},
\end{equation}
which results more complicate than the original one. Using the canonical
coordinates it is reduced the order in the following way i.e.
\begin{equation}
z^{\prime}=-\frac{z(1+zs+\left( Ks-2\right) s^{2}z^{2})}{s},  \label{abel}
\end{equation}
which is first-order ode type Abel, where $z=\frac{du}{ds}$,
\begin{equation}
s=\rho t^{2}\text{ \ \ \ and \ \ }z=\frac{1}{t^{2}\left(
\rho^{\prime}t+2\rho\right) }.
\end{equation}
The solution to eq. (\ref{abel}) is:
\begin{equation*}
C_{1}-\frac{\sqrt{-4z^{2}s^{2}+2Ks^{3}z^{2}+4zs-1}}{sz}+
\end{equation*}
\begin{equation}
+2\arctan\left( \frac{1-2zs}{\sqrt{-4z^{2}s^{2}+2Ks^{3}z^{2}+4zs-1}}\right)
=0.
\end{equation}

To end we will see another transformation. We are interesting into knowing
the transformation that produces the operator $X=t\partial_{t}+\rho%
\partial_{\rho }.$ Knowing that the canonical variables are obtained by
solving the equations $\left( Xs=0,Xu=1\right) $ we find that
\begin{equation}
s=\frac{\rho}{t}\text{ \ \ \ and \ \ \ \ }u(s)=\ln t,
\end{equation}
and therefore
\begin{equation}
t=e^{u(s)}\text{ \ \ and \ \ \ }\rho=se^{u(s)}
\end{equation}
with these new variables, equation (\ref{A_ber1}) simplifying yields
\begin{equation}
u^{\prime\prime}=-u^{\prime2}-\frac{u^{\prime}}{s}-\left(
1+Ke^{3u(s)}\right) su^{\prime3}
\end{equation}
which has a particular solution
\begin{equation}
u=\frac{1}{3}\ln\left( \frac{2}{Ks}\right)
\end{equation}
finding in this way that in the original variables it yields
\begin{equation}
\rho=\frac{2}{Kt^{2}}
\end{equation}
which is the solution obtained in section (\ref{sol1}).

Once again if one insist in solving equation (\ref{A_ber1}) through DA
(applying the Pi-theorem, see\cite{to5} for details) it is founded that \
with respect to the dimensional base $\frak{B}=\left\{ \rho,T\right\} $ each
quantity has the following dimensional equation $\left[ \rho\right] =\rho,%
\left[ t\right] =t$ and $\left[ K\right] =\rho^{-1}t^{-2}.$ Therefore, \ we
find in \ a trivial way that:
\begin{equation}
\begin{array}{r|rrr}
& \rho & K & t \\ \hline
\rho & 1 & -1 & 0 \\
T & 0 & -2 & 1
\end{array}
\Longrightarrow\rho\thickapprox\frac{1}{Kt^{2}}.
\end{equation}

It is observed that if for example we work with a new dimensional base like $%
\frak{B}=\left\{ L,M,T\right\} ,$ i.e. the usual one, the above quantities
have the following new dimensional equations: $\left[ \rho\right]
=L^{-1}MT^{-2},\left[ t\right] =T$ and $\left[ K\right] =LM^{-1}.$
Therefore, we have a new overdetermined system of equations which determine $%
\rho,$
\begin{equation}
\begin{array}{r|rrr}
& \rho & K & t \\ \hline
L & -1 & 1 & 0 \\
M & 1 & -1 & 0 \\
T & -2 & 0 & 1
\end{array}
\Longrightarrow\rho\thickapprox\frac{1}{Kt^{2}}.
\end{equation}

As it is observed we have obtained the same solution than the obtained one
in the previous sections, but now making only a simple dimensional
considerations.


\begin{thebibliography}{99}


\bibitem{K0_1}  X. Wang et al astro-ph/0105091.

\bibitem{K0_2}  C. B. Netterfield et l astro-ph/0104460

\bibitem{K0_3}  N. V. Halverson et al astro-ph/0104489

\bibitem{K0_4}  A. T. Lee et al astro-ph/0104459

\bibitem{Ma_1}  I. Prigogine and J. Geheniau, \textit{Proc. Nat. Acad. Sci.
USA} \textbf{83}, 6246 (1986).

\bibitem{Ma_2}  I. Prigogine, J. Geheniau, E. Gunzig and P. Nardone, \textit{%
Proc. Nat. Acad. Sci. USA} \textbf{83}, 7428 (1988).

\bibitem{Ma_3}  I. Prigogine, J. Geheniau, E. Gunzig and P. Nardone, \textit{%
Gen. Rel. Grav.} \textbf{21}, 767 (1989).

\bibitem{Ma_4}  J. A. S. Lima, A. S. M. Germano and L.R.W. Abramo, \textit{%
Phys.Rev.} \textbf{D53}, 4287 (1996).

\bibitem{ad1}  Birkhoff, G. ``\textit{Hydrodynamics. -A Study in Logic,
Facts and Similaritude}'' (1950) 1st Ed. Princeton.

\bibitem{ad2}  Sedov, L.I. ``\textit{Similarity and Dimensional Methods in
Mechanics}''. (1959) Infosearch Ltd. London.\

\bibitem{ad3}  J. Palacios. ``\textit{Dimensional Analysis}'' Mcmillan.
London 1964.

\bibitem{ad4}  K. Kurth. ``\textit{Dimensional Analysis and Group Theory in
Astrophysics}'', Pergamon Press, Oxford, 1972.

\bibitem{ad5}  G.I. Barenblatt. ``\textit{Similarity, self-similarity and
intermediate asymptotics}''. (1979) Consultants. Bureau New York. G.I.
Barenblatt. ``\textit{Scaling, Self-Similarity and Intermediate Asymptotics}%
'', Cambridge University Press, Cambridge, 1996.

\bibitem{ad6}  M. Casta\~{n}s. All this book.

\bibitem{to1}  J. A. Belinch\'{o}n, ``Standard Cosmology Through
Similarity'', physics/9811016.

\bibitem{to2}  J. A. Belinch\'{o}n \& A. Alfonso-Faus, \textit{Int. J. Mod.
Phys.} \textbf{D10}, 299 (2001).

\bibitem{to3}  J. A. Belinch\'{o}n, \textit{Int. J. Mod. Phys.} \textbf{D11 }%
, 527 (2002).

\bibitem{to4}  J. A. Belinch\'{o}n. \textit{Astro. Spac. Scien. }\textbf{281}%
, 765 (2002).

\bibitem{tobar}  E.M. Tobar. ``\textit{Global Description of the Fine
Structure Constant and its Variation Independent of Unit System}''.
hep-ph/0306230.

\bibitem{g1}  J. F. Cari\~{n}ena, and M. Santander, Advances in Electronics
and Electron Physics 72, 182 (1988).

\bibitem{g2}  G. W. Bluman and\ J. D. Cole. ``\textit{Similarity Methods
for Differentials Equations}''. (1974) Appl. Math. Sci. N13 Springer-Verlang
New York. G. W. Blumann and S.W. Kumei ``\textit{Symmetries and Differential Equations%
}''. Springer Verlang (1989). G. W. Blumann and S.C. Anco. ``\textit{Symmetries and Integration Methods for Differential Equations}". Springer Verlang (2002).

\bibitem{g3}  L. V. Ovsiannikov, \textit{Group Analysis of Differential
Equations} (Academic Press 1982).

\bibitem{g4}  N. H. Ibragimov. ``\textit{Transformation Groups Applied to
Mathematical Physics}''. D. Reidel (1985). N. H. Ibragimov. ``\textit{%
Elementary Lie Group Analysis and Ordinary Differential Equations''} (John
Wiley \& Sons, 1999).

\bibitem{g5}  R. Seshadri and\ T. Y. Na. ``\textit{Group Invariance in
Engineering Boundary}'' Value Problems. Springer-Verlang. NY, 1985.

\bibitem{g6}  H. Stephani, ``\textit{Differential Equations: Their Solutions
Using Symmetries''} (Cambridge University Press 1989).

\bibitem{g7}  P. T. Olver, ``\textit{Applications of Lie Groups to
Differential Equations''} (Springer-Verlang, 1993).

\bibitem{g8}  L. Dresner, ``\textit{Applications of Lie%
\'{}%
s Theory of Ordinry and Partial Differential Equations}''. IOP 1999

\bibitem{g9}  G. Bauman, ``\textit{Symmetry Analysis of Differential
Equations}''. Springer Verlang (2000).

\bibitem{g10}  P. E. Hydon, ``\textit{Symmetry Methods for Differential
Equations''} (Cambridge University Press, 2000).

\bibitem{g11}  B. J. Cantwell, ``\textit{Introduction to Symmetry Analysis''}
(Cambridge University Press, Cambridge, 2002).

\bibitem{pol1}  M. Szydlowki and M. Heller, \textit{Acta Physica Polonica}
\textbf{B14}, 571 (1983).

\bibitem{pol2}  M. Biesida, M. Szydlowki and T. Szczesny, \textit{Acta
Cosmologica}, Fasciculus XVI 115 (1989).

\bibitem{pol3}  C. B. Collins, \textit{J. Math. Phys. }\textbf{18}, 61374
(1977).

\bibitem{h1}  B. J. Carr and A. A. Coley, \textit{Class. Quantum Grav.}
\textbf{16}, R31 (1999).

\bibitem{h2}  G. I. Barenblatt and Y. B. Zeldovich, \textit{Ann. Rev. Fluid
Mech.} \textbf{4}, 285 (1972).

\bibitem{h3}  D. M. Eardley, \textit{Commun. Math. Phys.} \textbf{37}, 287
(1974).

\bibitem{h4}  J. Wainwright, \textit{Gen. Rel. Grav.} \textbf{16}, 657
(1984).

\bibitem{to5}  J. A. Belinch\'{o}n, ``\textit{Differential equations through Symmetries}". In preparation.
 Preprint: \url{http://www.matematicas.net}.



\end{thebibliography}
\end{document}